\documentclass[11pt]{article}

\usepackage[top = 30 mm, bottom = 30mm, left = 20mm, right=20mm]{geometry}
\usepackage{amsmath, amssymb}
\usepackage[english]{babel}
\usepackage[utf8]{inputenc}
\usepackage{latexsym}
\usepackage{graphicx}
\usepackage{lmodern}
\usepackage{jheppub}
\usepackage{slashed}
\usepackage{charter}
\usepackage{xcolor}
\usepackage{color}
\usepackage{float}
\usepackage{bm}
\usepackage{braket}

\numberwithin{equation}{section}

\newcommand{\be}{\begin{equation}}
\newcommand{\ee}{\end{equation}}
\newcommand{\bear}{\begin{array}}
\newcommand{\eear}{\end{array}}

\title{Massive gravitino scattering amplitudes and\\the unitarity cutoff of the new Fayet-Iliopoulos terms}
\author[a, b]{Ignatios Antoniadis}
\author[a]{Anthony Guillen}
\author[a, c]{François Rondeau}

\affiliation[a]{Laboratoire de Physique Th\'eorique et Hautes Energies - LPTHE\\
Sorbonne Universit\'e, CNRS, 4 Place Jussieu, 75005 Paris, France}
\affiliation[b]{Department of Mathematical Sciences, University of Liverpool\\
Liverpool L69 7ZL, United Kingdom}
\affiliation[c]{Department of Physics, University of Cyprus, Nicosia 1678, Cyprus}

\emailAdd{antoniad@lpthe.jussieu.fr}
\emailAdd{aguillen@lpthe.jussieu.fr}
\emailAdd{frondeau@lpthe.jussieu.fr}

\date{} 

\abstract{We compute the $2\rightarrow 2$ gravitino scattering amplitudes at tree level in supergravity theories where supersymmetry is spontaneously broken. In the unitary gauge, the gravitino becomes massive (of mass $m_{3/2}$) by absorbing the Goldstino, and the scattering amplitudes of its longitudinal polarisations grow with energy as $\kappa^2 E^4/m_{3/2}^2$, signaling a potential breakdown of unitarity at a scale $\Lambda^2\sim m_{3/2}/\kappa\sim M_{\mathrm{SUSY}}^2$. As we show explicitly in the Polonyi model, this leading term is cancelled by the contributions coming from the scalar partner of the Goldstino (sgoldstino), restoring perturbative unitarity up to the Planck scale. This is expected since supersymmetry is spontaneously broken, in analogy with the situation occuring in the Standard Model, where massive gauge bosons scattering preserves unitarity at high energy once we consider the contributions from the Higgs boson. However, when supersymmetry is broken by the new Fayet-Iliopoulos $D$-term, with ungauged R-symmetry, the above cancellation does not occur. In this case, the unbroken phase is singular and there is no contribution able to cancel the quartic divergences of the amplitudes, leading to a cutoff $\Lambda\sim M_\mathrm{SUSY}$ where the effective theory breaks down. The same behaviour is obtained when supersymmetry is non-linearly realised.}

\begin{document}

\maketitle\newpage

\section{Introduction}

\vfill

Supergravity is a very interesting theory, deeply rooted in the landscape of theoretical physics; it can notably be viewed as the low-energy, field-theoretical description of superstring theory. If supergravity is indeed realised in nature, the supersymmetry on which it is based must break at some point to recover the non-supersymmetric Standard Model at lower energies. This (spontaneous) breaking can either be sourced by vacuum expectation values (vevs) of auxiliary $F$-terms, associated with chiral multiplets, or by vevs of auxiliary $D$-terms, associated with vector multiplets. A simple prototype of $F$-term supergravity breaking, using a linear superpotential, is due to Polonyi \cite{Polonyi:1977pj}. In global supersymmetry, $D$-breaking is obtained with the Fayet-Iliopoulos (FI) term \cite{Fayet:1974jb}. Writing this in supergravity requires gauging the R-symmetry \cite{Freedman:1976uk}, with various consequences. For instance, in the absence of chiral multiplets, it forbids an explicit mass term for the gravitino, because the superpotential should then transform under R-symmetry \cite{Barbieri:1982ac}, along with other complications that blur the connection with superstring theory, unless the FI term is field dependent \cite{Komargodski:2009pc, Dienes:2009td}.

\vfill

More recently, a new embedding of the FI term in supergravity has been found that does not require gauging the R-symmetry \cite{Cribiori:2017laj, Cribiori:2018dlc}. It was then noticed that when coupled with chiral multiplets, the simplest generalisation of the original new FI term breaks invariance under Kähler transformations \cite{Cribiori:2018dlc, Antoniadis:2018cpq}, so a Kähler invariant construction was proposed in \cite{Antoniadis:2018oeh} and further generalised in \cite{Antoniadis:2019nwz}. These new FI terms have also found cosmological applications, to uplift the cosmological constant \cite{Antoniadis:2018oeh} or to construct models of inflation \cite{Antoniadis:2018cpq, Antoniadis:2019nwz, Jang:2021fce, Jang:2022sql}. A peculiarity of these terms is that they contain inverse powers of the auxiliary field $D$, which makes them singular when supersymmetry is unbroken. In addition, they include a plethora of non-renormalisable and multi-fermion interactions, which in general break the validity of the effective theory above a certain ultraviolet (UV) cutoff. This cutoff was estimated by dimensional analysis in \cite{Jang:2021vpb} and with more direct arguments in \cite{Jang:2022sql}; it was found to sit at the supersymmetry breaking scale.

\vfill

There is however a limitation to the dimensional analysis performed in \cite{Jang:2022sql, Jang:2021vpb}: the cutoff obtained by inspection of the interactions at the level of the Lagrangian may be uplifted due to cancellations in the corresponding scattering amplitudes. Let us consider, for example, the scattering of four massive gauge bosons with longitudinal polarisation in the Standard Model. Each of the external polarisations scale like $E/v$, where $E$ is the energy of the process and $v$ is the vacuum expectation value of the Higgs field, while there is a quartic interaction that does not depend on energy; so we expect the $2\rightarrow2$ amplitude to behave as $E^4/v^4$, breaking perturbative unitarity around a cutoff $\Lambda_\mathrm{cut}\sim v$. However, it is well known that cancellations between the pure gauge sector and the Higgs boson channels lead to an amplitude that does not grow with energy and respects perturbative unitarity; this is one of the interests of the Higgs mechanism. These cancellations are not miraculous, but they happen because of the relations that link the gauge and Higgs couplings upon spontaneous symmetry breaking. 

\vfill\newpage\vfill

In this paper, we extend this observation to models of spontaneously broken supergravity. In this case, the analog of massive gauge bosons are massive gravitinos. After short reminders on perturbative unitarity and Standard Model gauge boson scattering in Section 2, we detail the computation of longitudinal gravitinos scattering amplitude in Section 3 taking into account at first only gravitational interactions. We find that the naive scaling $\mathcal{M}\sim\kappa^2E^6/m_{3/2}^4$ that one could expect from the Lagrangian is modified to $\mathcal{M}\sim\kappa^2E^4/m_{3/2}^2$ by a first cancellation, which can be explained using the Goldstino equivalence theorem. In Section 4, we consider the standard Polonyi model of supergravity breaking. In this case, the scalar of the chiral multiplet, which is the sgoldstino, contributes to the amplitudes, to precisely cancel the leading terms of Section 3, leaving amplitudes that scale at most like $\mathcal{M}\sim\kappa^2E^2$, and a cutoff at the Planck scale. This second cancellation can be understood, as in the Standard Model, from the fact that the gravitino-graviton and the gravitino-scalar couplings are related to ensure supersymmetry. An intuition behind it is that the microscopic theory also describes the unbroken phase, where gauge bosons / gravitinos are massless and their amplitudes are well-behaved with respect to perturbative unitarity. Therefore, it is natural to expect the same behaviour to extend in the broken phase. When supergravity is broken by the new FI term, however, as we mentioned earlier the unbroken phase is singular and thus it is not clear to expect a cancellation of the terms $\mathcal{M}\sim\kappa^2E^4/m_{3/2}^2$ from the gravitational sector. Indeed, in Section 5, after reviewing the construction of the FI term and the power counting argument of \cite{Jang:2021vpb}, we show that this cancellation cannot occur, because the new FI term does not contain any interaction with two gravitinos and the vector boson of the goldstino multiplet, the only extra field that could contribute to the $2\rightarrow2$ scattering at tree level. This leaves a cutoff at the supersymmetry breaking scale $\Lambda_\mathrm{cut}\sim M_\mathrm{SUSY}$, confirming the result of \cite{Jang:2022sql, Jang:2021vpb}.  

\vfill\newpage\vfill

\section{Unitarity and the Higgs mechanism}
\label{sect:Higgs}

Unitarity is an important property that must be satisfied by any consistent Quantum Field Theory. If it was not, it could mean that some processes occur with probability greater than one, which is hard to accept. Imposing unitarity of the S-matrix leads to the so-called optical theorem, implying in particular that scattering amplitudes cannot be arbitrarily large. At first approximation, we can apply the following bound to the tree-level scattering amplitudes 
\begin{equation}\label{eq:unitarity-bound}
\mathrm{Im}\mathcal{M} \leq |\mathcal{M}|^2 \quad\rightarrow\quad
|\mathcal{M}| < 1.
\end{equation}

Strictly speaking, unitarity is not necessarily violated above this bound; it could also signal that new states should be included in the theory, or a breakdown of perturbation theory. In any case, this is a point where something happens. In the Standard Model, it turns out that the Higgs boson plays a crucial role in maintaining perturbative unitarity after electroweak symmetry breaking. Let us briefly review this phenomenon before exploring how it extends to models of supergravity breaking in the next Sections. The problem is the following: when electroweak symmetry is broken, the $W^\pm$ and $Z$ bosons become massive and can propagate with a longitudinal polarisation. But longitudinally polarised vectors scale like $E/v$ and bring factors of $E$ in scattering amplitudes (\ref{eq:longitudinal-polarization}), where $E$ is the energy of the process and $v$ is the vacuum expectation value of the Higgs field associated to the Fermi constant. Consider for instance $W^+Z \rightarrow W^+Z$. If we ignore the physical Higgs boson, there are three graphs contributing to this amplitude: two for $s$ and $u$ channel with a $W$ exchange, and one coming from the quartic interaction between the gauge bosons. The cubic vertex comes with a factor of $E$, the $W$ propagator with a factor $1/E^2$, and the quartic vertex has no $E$ dependence. Therefore, the amplitude with four longitudinal polarisations could scale like $E^4/v^4$ and break (\ref{eq:unitarity-bound}) when the scattering energy approaches $v$. Computing these three graphs more precisely yields \cite{Schwartz:2014sze}
\begin{equation}\label{eq:amplitude-gauge}
\mathcal{M}_{\mathrm{gauge}}(W_LZ_L\rightarrow W_LZ_L) = \frac{t}{v^2} + \mathcal{O}(1),
\end{equation}

where $t$ is a Mandelstam variable, $t = (p_1 - p_3)^2 \sim E^2$. Instead of $E^4/v^4$, this amplitude scales like $E^2/v^2$. In fact, its three contributions do separately have terms scaling like $E^4/v^4$, but they cancel each other in the sum. This first cancellation can be explained using the Goldstone theorem, since the corresponding Goldstone amplitudes scale at most like $E^2/v^2$ at tree level. A second cancellation occurs when we take into account the physical Higgs boson, which can be exchanged in the $t$ channel, adding another contribution to the amplitude that cancels the first term of (\ref{eq:amplitude-gauge})
\begin{equation}
\mathcal{M}_{h}(W_LZ_L\rightarrow W_LZ_L) = -\frac{t}{v^2} + \mathcal{O}(1).
\end{equation}

The result of these cancellations is that, with the Higgs boson, perturbative unitarity is maintained after electroweak symmetry breaking. This seems reasonable: in the unbroken phase, gauge bosons are massless and their amplitudes are well-behaved. So there is no reason to expect a difference in the broken phase, which is described by the same underlying theory.

\vfill\newpage\vfill

\section{Massive gravitino scattering}
\label{sect:mass_grav_scatt}
In this Section, we compute the massive gravitino scattering amplitudes $\psi\psi\rightarrow\psi\psi$ in $\mathcal{N}=1$ supergravity. In particular, we compute these amplitudes with external helicities $\pm 1/2$. Our purpose is to compare their behaviour with the case of longitudinal massive gauge boson amplitudes, and in particular to observe if the same kind of cancellations occur, as described previously. Our starting point is the following action \cite{Freedman:2012zz, Cremmer:1982en}
\begin{equation}\label{pure_sugra_action}
\mathcal{S} = \int d^4x e\left(\frac{1}{2\kappa^2}R(e) - \frac{1}{2}\bar\psi_\mu\gamma^{\mu\nu\rho}\left(\partial_\mu + \frac{1}{4}\omega_\nu^{ab}(e)\gamma_{ab}\right)\psi_\rho + \frac{1}{2}m_{3/2}\bar\psi_\mu\gamma^{\mu\nu}\psi_\nu + \mathcal{L}_{4f}\right),\\
\end{equation}
where
\begin{equation}\label{gravitino_quartic_interaction}
\mathcal{L}_{4f} = 
-\frac{\kappa^2}{32}(\bar\psi^\mu\gamma^\rho\psi^\nu)(\bar\psi_\mu\gamma_\rho\psi_\nu)
-\frac{\kappa^2}{16}(\bar\psi^\rho\gamma^\mu\psi^\nu)(\bar\psi_\rho\gamma_\nu\psi_\mu)
+\frac{\kappa^2}{8}(\bar\psi^\rho\gamma^\mu\psi_\mu)(\bar\psi_\rho\gamma^\nu\psi_\nu).
\end{equation}

Here, $R(e)$ is the usual curvature scalar, and ${\omega_\mu}^{ab}(e)$ is the spin connection, given in terms of the frame field $e^a_\mu$ by
\begin{equation}\label{eq:spin-connection}
{\omega_\mu}^{ab}(e) = 2e^{\nu[a}\partial_{[\mu}{e_{\nu]}}^{b]} - e^{\nu[a}e^{b]^\sigma}e_{\mu c}\partial_\nu{e_\sigma}^c.
\end{equation}

In all the expressions above, latin letters $a, b, c, ...$ are used for the 'flat' indices and the greek letters $\mu, \nu, \rho, ...$ for the 'curved' ones. The action (\ref{pure_sugra_action}) is not supersymmetric, because the gravitino mass term has been added by hand; it is of the form $\mathcal{L} = \mathcal{L}_{\mathrm{grav}} + \mathcal{L}_{m_{3/2}}$, where $\mathcal{L}_{\mathrm{grav}}$ is the Lagrangian of pure $\mathcal{N}=1$ supergravity. In the next Sections, we will consider models where $\mathcal{L}_{m_{3/2}}$ is generated by spontaneous supersymmetry breaking. These models allow for a vanishing cosmological constant, so we choose to work in a Minkowski background.

\vfill

We consider that the gravitino is a Majorana fermion. Therefore, there are four graphs contributing to the $\psi\psi\rightarrow\psi\psi$ scattering amplitude at tree level: one for each of the $s, t, u$ channels, with a graviton in the middle, and one coming from the quartic interaction (\ref{gravitino_quartic_interaction}). Three ingredients are required to compute the $s, t, u$ graphs: the propagator of the graviton, the $h\psi^2$ vertex, and the external polarisation of the gravitinos. Let us obtain all these ingredients.

\vfill

The graviton $h$ is the fluctuation of the metric around a background, Minkowski in our case
\begin{equation}\label{expansion_metric}
g_{\mu\nu} = \eta_{\mu\nu} + 2\kappa h_{\mu\nu}.
\end{equation}

Its propagator is obtained from a standard procedure: expand the Einsten-Hilbert part of the action to second order in $h$, read the kinetic operator of the graviton, supplement the action with a gauge fixing term, in order to make this operator invertible, and compute its inverse. If we make the standard choice $\mathcal{L}_{g.f.} = -(\partial^\mu h_{\mu\nu} - 1/2\partial_\nu h)^2$, we obtain as propagator
\begin{equation}\label{graviton_propagator}
P^{\alpha\beta}_{\gamma\delta}(k) = -\frac{i}{2k^2}(\delta^\alpha_\gamma\delta^\beta_\delta + \delta^\beta_\gamma\delta^\alpha_\delta - \eta^{\alpha\beta}\eta_{\gamma\delta}).
\end{equation}

\vfill\newpage\vfill

Let us now compute the $h\psi^2$ vertex. First, note that in terms of the frame field
\begin{equation}\label{expansion_frame_field}
e^a_\mu = \delta^a_\mu + \kappa h^a_\mu \quad\text{and}\quad
e^\mu_a = \delta^\mu_a - \kappa h^\mu_a,
\end{equation}

and after this expansion is made, there is no distinction between flat and curved indices, since we work around a flat background. Before developing (\ref{pure_sugra_action}) to obtain its $h\psi^2$ part, it is convenient to write:
\begin{equation}
-\frac{1}{2}e\bar\psi_\mu\gamma^{\mu\nu\rho}D_\nu\psi_\rho = \frac{i}{2}\varepsilon^{\mu\nu\rho\sigma}\bar\psi_\mu\gamma_*\gamma_\sigma D_\nu\psi_\rho,
\end{equation}

where $D_\nu$ is the covariant derivative that can be read in (\ref{pure_sugra_action}), $\varepsilon^{\mu\nu\rho\sigma}$ is the Levi-Civita form and $\gamma_*$ is the fifth gamma matrix; our conventions are taken from \cite{Freedman:2012zz}. This rewriting is convenient, because several factors of the frame field in $e\gamma^{\mu\nu\rho}=ee^\mu_ae^\nu_be^\rho_c\gamma^{abc}$ are replaced by only one in $\gamma_\sigma=e_\sigma^a\gamma_a$. The frame field also appears in $\omega_\mu^{ab}(e)$, which is at first order in $h$
\begin{equation}
\omega_\mu^{ab}(e) = -\kappa\partial^{[a} {h^{b]}}_\mu.
\end{equation}

Using all this information, we can easily write the $h\psi^2$ part of the Lagrangian in (\ref{pure_sugra_action})
\begin{eqnarray}
\mathcal{L}_{h\psi^2} = \frac{i\kappa}{2}\varepsilon^{\mu\nu\rho\sigma}h^a_\sigma\bar\psi_\mu\gamma_*\gamma_a\partial_\nu\psi_\rho
- \frac{i\kappa}{4}\varepsilon^{\mu\nu\rho\sigma}\partial^a h^b_\nu\bar\psi_\mu\gamma_*\gamma_\sigma\gamma_{ab}\psi_\rho\\
+ \frac{\kappa}{2}m_{3/2}h\bar\psi_\mu\gamma^{\mu\nu}\psi_\nu - \kappa m_{3/2}h^\mu_\rho
\bar\psi_\mu\gamma^{\rho\nu}\psi_\nu.\label{graviton_gravitino_interaction_1}
\end{eqnarray}

The first line can be simplified. Using standard gamma-matrix manipulations, one can show that $\varepsilon^{\mu\nu\rho\sigma}\gamma_*\gamma_a = 4i\gamma^{[\mu\nu\rho}{\delta^{\sigma]}}_a$ and $\varepsilon^{\mu\nu\rho\sigma}\gamma_*\gamma_\sigma\gamma_{ab} = 6i{\gamma^{[\mu\nu}}_{[b}{\delta^{\rho]}}_{a]} + 6i\gamma^{[\mu}{\delta^\nu}_{[b}{\delta^{\rho]}}_{a]}$, leading to
\begin{equation}\label{graviton_gravitino_interaction_2}
\frac{i\kappa}{2}\varepsilon^{\mu\nu\rho\sigma}h^a_\sigma\bar\psi_\mu\gamma_*\gamma_a\partial_\nu\psi_\rho
= -\frac{\kappa}{2}h_{\alpha\beta}\bar\psi_\mu(\eta^{\alpha\beta}\gamma^{\mu\rho\nu}-\eta^{\alpha\mu}\gamma^{\beta\rho\nu}+\eta^{\alpha\rho}\gamma^{\beta\mu\nu}-\eta^{\alpha\nu}\gamma^{\beta\mu\rho})\partial_\rho\psi_\nu, 
\end{equation}

and
\begin{equation}\label{graviton_gravitino_interaction_3}
- \frac{i\kappa}{4}\varepsilon^{\mu\nu\rho\sigma}\partial^a h^b_\nu\bar\psi_\mu\gamma_*\gamma_\sigma\gamma_{ab}\psi_\rho = -\frac{\kappa}{2}\partial^\rho h_{\alpha\beta}\bar\psi_\mu(\delta^\mu_\rho\eta^{\alpha\beta}\gamma^\nu
- \delta^\alpha_\rho\eta^{\beta\mu}\gamma^\nu - \delta_\rho^\mu\eta^{\alpha\nu}\gamma^\beta)\psi_\nu.
\end{equation}

Combining (\ref{graviton_gravitino_interaction_1}), (\ref{graviton_gravitino_interaction_2}), (\ref{graviton_gravitino_interaction_3}), where we took care to write the indices as $h_{\alpha\beta}\bar\psi_\mu\cdots\psi_\nu$, we finally write the vertex. We consider all the momenta incoming and label them as $h(k)\bar\psi(p)\psi(q)$; we also denote $t = p-q$. In principle, the vertex should be symmetrised with respects to $(\alpha\beta)$, but since it is always contracted with the graviton propagator, which is also symmetric, we keep only one side of the $(\alpha\beta)$ symmetry in the vertex. We also take into account that the two gravitinos are Majorana and can be interchanged, with a sign that depends on the rank of the gamma matrix inserted between them (see \cite{Freedman:2012zz}). Finally, there is a factor of $-i$ associated with each derivative, along with the usual global factor of $i$
\begin{eqnarray}\label{eq:graviton-gravitino-vertex}
V^{\mu, \nu, \alpha\beta}_{h\psi^2}(t, k) &=& \kappa/2(\eta^{\alpha\beta}\gamma^\mu k^\nu 
+ \eta^{\alpha\mu}\gamma^\nu k^\beta - \eta^{\alpha\mu}\gamma^\beta k^\nu+\eta^{\alpha\beta}\gamma^{\mu\rho\nu}t_\rho+\eta^{\alpha\mu}\gamma^{\beta\nu\rho}t_\rho \nonumber\\
&&\phantom{-}-\eta^{\alpha\beta}\gamma^\nu k^\mu - \eta^{\alpha\nu}\gamma^\mu k^\beta + \eta^{\alpha\nu}\gamma^\beta k^\mu
+ \gamma^{\alpha\mu\nu}t^\beta- \eta^{\alpha\nu}\gamma^{\beta\mu\rho}t_\rho)\nonumber\\
&+& i\kappa m_{3/2}(\eta^{\alpha\beta}\gamma^{\mu\nu} - \eta^{\alpha\mu}\gamma^{\beta\nu} + \eta^{\alpha\nu}\gamma^{\beta\mu}).
\end{eqnarray}

\vfill\newpage\vfill

Since in our computation the gravitinos are external and on-shell, we can simplify the vertex using the equations of motion of the free gravitinos; for instance if we use $\gamma\cdot\psi = 0$, then
\begin{eqnarray}\label{graviton_gravitino_vertex_simp}
V^{\mu, \nu, \alpha\beta}_{h\psi^2}(t, k) &\sim& \kappa/2(- \eta^{\alpha\mu}\gamma^\beta k^\nu+ \eta^{\alpha\beta}\eta^{\mu\nu}\slashed t - \eta^{\alpha\mu}\gamma^\beta t^\nu + \eta^{\alpha\mu}\eta^{\nu\beta} \slashed t \nonumber\\
&&\phantom{--}\eta^{\alpha\nu}\gamma^\beta k^\mu -\eta^{\mu\nu}\gamma^\alpha t^\beta+ \eta^{\alpha\nu}\gamma^\beta t^\mu - \eta^{\alpha\nu}\eta^{\mu\beta}\slashed t)\nonumber\\
&+& i\kappa m_{3/2}(\eta^{\alpha\mu}\eta^{\beta\nu} + \eta^{\alpha\nu}\eta^{\beta\mu} - \eta^{\alpha\beta}\eta^{\mu\nu}).
\end{eqnarray}

We could also use $p\cdot\psi = 0$ and $(i\slashed p - m_{3/2})\psi_\mu=0$, but the former does not bring much simplification, and using the latter requires knowing if the external leg connected to the vertex is ingoing or outgoing, in order to fix the sign of the momentum, which is not convenient. \footnote{Our vertex differs from \cite{Bjerrum-Bohr:2010eeh}, where however the terms with $\gamma^\mu$ do not  vanish using $\gamma\cdot\psi=0$ as they should.}

\vfill

Now that we have the graviton propagator and the $h\psi^2$ vertex, the last thing we need are the external polarisations of the gravitinos. These are vector-spinors, solutions of the free Rarita-Schwinger equation, which can be recasted in momentum space as \cite{Moroi:1995fs}
\begin{equation}\label{equation_motion_gravitino}
\gamma\cdot\psi^\lambda(p) = 0, \quad p\cdot\psi^\lambda(p) = 0 \quad\text{and}
\quad (i\slashed p - m_{3/2})\psi_\mu^\lambda(p) = 0.
\end{equation}

The solutions can be written in a simple way using Clebsch-Gordon decomposition
\begin{eqnarray}\label{external_polarizations_gravitino}
\psi_\mu^{++} = \varepsilon_\mu^+u^+, \quad
\psi_\mu^+ = \sqrt{2/3}\varepsilon^0_\mu u^+ + \sqrt{1/3}\varepsilon_\mu^+u^-\nonumber\\
\psi_\mu^{--} = \varepsilon_\mu^-u^-, \quad
\psi_\mu^- = \sqrt{2/3}\varepsilon^0_\mu u^- + \sqrt{1/3}\varepsilon_\mu^-u^+,
\end{eqnarray}

where the $\varepsilon^{\pm, 0}_\mu$ are the standard polarisation vectors, transverse and mutually orthogonal, while the spinors $u^\pm$ are solution of the free Dirac equation. The Dirac equation has two more solutions, usually called $v^\pm$, which can replace $u^\pm$ in the previous equation. The choice between the two for any external leg is done in the same way as for spin-$1/2$ fermions
\begin{itemize}
\item[$\cdot$]$u$ for an ingoing ``gravitino'', $\bar v$ for an ingoing ``anti-gravitino''
\item[$\cdot$]$\bar u$ for an outgoing ``gravitino'', $v$ for an outgoing ``anti-gravitino''.
\end{itemize}

Here, the bar stands for the Dirac conjugate $\bar\lambda^D = \lambda^\dagger\gamma^0$, while it was standing for the Majorana conjugate $\bar\lambda^M = \lambda^T C$ up to now. Note that the Dirac conjugate of a $v$-spinor is the Majorana conjugate of the corresponding $u$-spinor since they are charge conjugates $v^\pm = (u^\pm)^c$.

\vfill
 
Since the gravitinos are Majorana, we cannot talk about ``gravitinos'' and ``anti-gravitinos'' in the usual sense. In the present context, we simply use ``gravitino'' when the fermionic current attached to the leg is oriented towards the right of the graph, and ``anti-gravitino'' when it is oriented towards the left. The direction of this current on each fermionic line is arbitrary, as long as it remains consistent, to form bilinears etc. The $h\psi^2$ vertex in (\ref{graviton_gravitino_vertex_simp}) has been (anti-)symmetrised in a way that takes into account this arbitrariness.

\vfill\newpage\vfill

We now give expressions for the $\varepsilon^{\pm, 0}$ and the $u^\pm, v^\pm$. For a momentum parameterised as
\begin{equation}
p^\mu = (p^0, |\bm p|\sin\theta\cos\varphi, |\bm p|\sin\theta\sin\varphi, |\bm p|\cos\theta),
\end{equation}

the $\varepsilon^{\pm, 0}$ are given by, up to a sign that can be adjusted for $\psi_\mu^\pm$ to satisfy the e.o.m. (\ref{equation_motion_gravitino})

\begin{equation}
\varepsilon^\mu_\pm = 1/\sqrt{2}(0, \cos\theta\cos\varphi\mp i\sin\varphi, \cos\theta\sin\varphi\pm i \cos\varphi, -\sin\theta),
\end{equation}

and
\begin{equation}\label{eq:longitudinal-polarization}
\varepsilon^\mu_0 = 1/m_{3/2}\left(|\bm p|, p^0\bm p/|\bm p|\right).
\end{equation}

For the $u^\pm$, we need to choose a representation of the $\gamma$-matrices; we used the Weyl one
\begin{equation}
\gamma^\mu = \left(\begin{matrix}0 & \sigma^\mu \\\bar\sigma^\mu & 0\end{matrix}\right) \quad\mathrm{where}\quad \sigma^\mu = (1, \sigma^i), \quad\bar\sigma^\mu = (-1, \sigma^i),
\end{equation}

and the $\sigma^i$ are the usual Pauli matrices. In the following, when computing the amplitude, we take as convention that all gravitinos are incoming. By energy conservation, some of them will then have $p^0 > 0$ and others $p^0 < 0$, and the expressions of $u^\pm$ differ in these two cases
\begin{eqnarray}\label{eq:external-spinors}
u^\pm(p) &=& \left(\begin{matrix}\phantom{-i}\sqrt{-\sigma\cdot p}\ \xi^\pm \\ \phantom{-}i\sqrt{\phantom{-}\bar\sigma\cdot p}\ \xi^\pm\end{matrix}\right) \quad\text{if}\quad p^0 > 0,\\
u^\pm(p) &=& \left(\begin{matrix}\phantom{-i}\sqrt{\phantom{-}\sigma\cdot p}\ \xi^\mp \\ -i\sqrt{-\bar\sigma\cdot p}\ \xi^\mp\end{matrix}\right)\quad\text{if}\quad p^0 < 0,
\end{eqnarray}

where
\begin{equation}
\xi^+ = \left(\begin{matrix}\cos(\theta/2) \\ \sin(\theta/2)\end{matrix}\right)
\quad\text{and}\quad
\xi^- = \left(\begin{matrix}-\sin(\theta/2)\\ \phantom{-}\cos(\theta/2)\end{matrix}\right).
\end{equation}

As mentioned before, the $v^\pm$ spinors are simply obtained as the charge conjugates of the $u^\pm$.

\vfill

We now have all the necessary ingredients to compute the $s, t, u$ graphs contributing to the $\psi\psi\rightarrow\psi\psi$ amplitudes: the propagator (\ref{graviton_propagator}), the vertex (\ref{graviton_gravitino_vertex_simp}) and the polarisations (\ref{external_polarizations_gravitino}). We just have to understand how to combine these. Drawing the graphs and writing the corresponding subamplitudes is  straightforward; the only question is: what are the relative signs between them? Before answering this, we give our conventions for the orientation of the fermionic currents with the following figure. Note that we consider all momenta ingoing.

\begin{figure}[ht]
    \centering
    \includegraphics[scale=0.50]{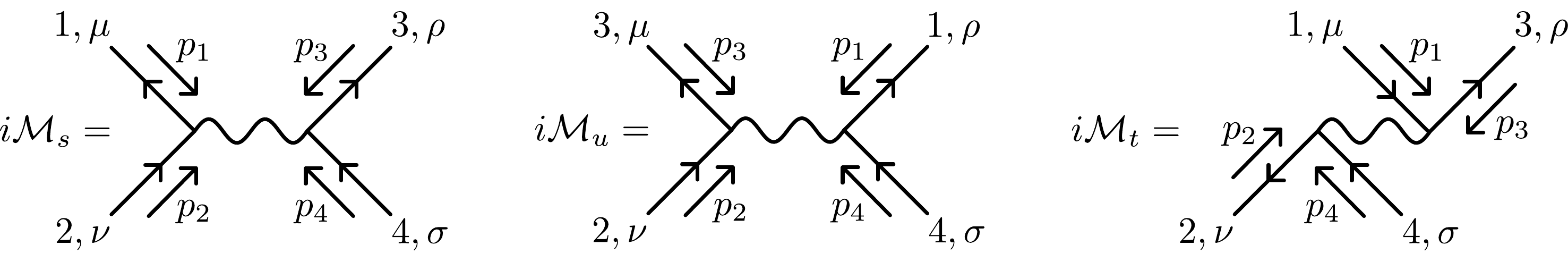}
    \caption{Conventions for the orientation of the fermionic lines in the $s,t,u$ channels.}
    \label{fig:diagrams}
\end{figure}

\vfill\newpage\vfill

Without loss of generality, we assume that the $s$-channel graph comes with a $+$ sign. From there, the $u$-channel is obtained by exchanging gravitinos $1$ and $3$, so it comes with a $-$. To obtain the $t$-channel, we invert the fermionic current around the first vertex before exchanging $1$ and $4$. Since inverting the fermionic current corresponds to exchanging $1$ and $2$, obtaining the $t$-channel amounts to two fermions exchanges, therefore it comes with a $+$ sign. These relative signs ensure that the total amplitude gets a minus sign under the exchange of any pair of gravitinos. This can be checked using the symmetries of the propagator (\ref{graviton_propagator}) and the vertex  (\ref{graviton_gravitino_vertex_simp}), but this is easier to see in the contribution of the quartic interaction (\ref{gravitino_quartic_interaction}). We write this contribution using the same principles as for the vertex (\ref{graviton_gravitino_vertex_simp}) and the $s, t, u$ graphs, to take into account that the four gravitinos connected to the vertex are fully equivalent
\begin{eqnarray}\label{quartic_amplitude_gravitino}
\mathcal{M}_{4f} = -\frac{1}{16}(\eta^{\alpha\beta}\eta^{\mu\rho}\eta^{\nu\sigma} + 2\eta^{\alpha\sigma}\eta^ {\beta\nu}\eta^{\mu\rho} - 4\eta^{\alpha\nu}\eta^{\beta\sigma}\eta^{\mu\rho} + [\mu\leftrightarrow\nu]+ [\rho\leftrightarrow\sigma])\nonumber\\
\times((\bar\psi_\mu^1\gamma_\alpha\psi_\nu^2)(\bar\psi_\rho^3\gamma_\beta\psi_\sigma^4)
-(\bar\psi_\mu^3\gamma_\alpha\psi_\nu^2)(\bar\psi_\rho^1\gamma_\beta\psi_\sigma^4)
+(\bar\psi_\mu^2\gamma_\alpha\psi_\nu^4)(\bar\psi_\rho^3\gamma_\beta\psi_\sigma^1).
\end{eqnarray}

Here, $[\mu\leftrightarrow\nu]$ and $[\rho\leftrightarrow\sigma]$ stands for the same term with antisymmetrised indices. Indeed, if we exchange for instance the gravitinos $1$ and $2$ in the previous formula, we obtain
\begin{eqnarray}\label{exchange_spinor}
(\bar\psi_\mu^1\gamma_\alpha\psi_\nu^2)(\bar\psi_\rho^3\gamma_\beta\psi_\sigma^4)&\underset{1\leftrightarrow2}{\rightarrow}&
(\bar\psi_\mu^2\gamma_\alpha\psi_\nu^1)(\bar\psi_\rho^3\gamma_\beta\psi_\sigma^4)\sim
-(\bar\psi_\mu^1\gamma_\alpha\psi_\nu^2)(\bar\psi_\rho^3\gamma_\beta\psi_\sigma^4)\nonumber\\
(\bar\psi_\mu^3\gamma_\alpha\psi_\nu^2)(\bar\psi_\rho^1\gamma_\beta\psi_\sigma^4)&\underset{1\leftrightarrow2}{\rightarrow}&
(\bar\psi_\mu^3\gamma_\alpha\psi_\nu^1)(\bar\psi_\rho^2\gamma_\beta\psi_\sigma^4)\sim
\phantom{-}(\bar\psi_\mu^2\gamma_\alpha\psi_\nu^4)(\bar\psi_\rho^3\gamma_\beta\psi_\sigma^1)\nonumber\\
(\bar\psi_\mu^2\gamma_\alpha\psi_\nu^4)(\bar\psi_\rho^3\gamma_\beta\psi_\sigma^1)&\underset{1\leftrightarrow2}{\rightarrow}&
(\bar\psi_\mu^1\gamma_\alpha\psi_\nu^4)(\bar\psi_\rho^3\gamma_\beta\psi_\sigma^2)\sim
\phantom{-}(\bar\psi_\mu^3\gamma_\alpha\psi_\nu^2)(\bar\psi_\rho^1\gamma_\beta\psi_\sigma^4),
\end{eqnarray}

where the $\sim$ means up to the symmetry with respects to $\mu\leftrightarrow\rho$ and $(\alpha\sigma)\leftrightarrow(\beta\nu)$ and antisymmetry with respects to $\mu\leftrightarrow\nu$ that is provided by the first parenthesis of (\ref{quartic_amplitude_gravitino}). Note that to obtain the first line we use the Majorana flip $\bar\psi_\mu^2\gamma_\alpha\psi_\nu^1 = \bar\psi_\nu^1\gamma_\alpha\psi_\mu^2$, where the usual minus sign is absent because the external polarisations are commuting spinors (in the plane wave expansion of a spinor, its anticommuting nature is carried by the operator coefficients that multiply the polarisations). Equations (\ref{exchange_spinor}) show that $i\mathcal{M}_4$ gets a minus sign under the exchange of gravitinos $1$ and $2$; one can show that this is true for any pair, as well as for the total amplitude including the $s, t, u$ channels, provided we sum them as, in our conventions
\begin{eqnarray}
\mathcal{M}_{\mathrm{grav}} = \mathcal{M}_s-\mathcal{M}_t+\mathcal{M}_u+\mathcal{M}_{4f}.
\end{eqnarray}

Finally, we have everything we need to compute the amplitudes. With four choices for the helicity of each external leg, there are $256$ possibilities, halved to $128$ by CPT symmetry. But we are mostly interested in the ones with external helicities $\pm 1/2$, to observe the cancellations of the most divergent terms with respect to the energy as in the case of longitudinal massive gauge boson amplitudes described in Section \ref{sect:Higgs}. This still gives $8$ possibilities after CPT, but for simplicity, and since the others behave in a similar way, here we will not report all of them.

\vfill

We write the amplitudes as a function of $E$ and $\theta$, where $E$ is the energy of the gravitinos $1$ and $2$ in the rest frame (so center of mass energy is $2E$) and $\theta$ is the angle between $1$ and $3$.

\vfill\newpage\vfill

The amplitudes with external helicities $\pm 1/2$ scale at most like $\kappa^2E^6/m_{3/2}^4$, with four factors $E/m_{3/2}$ from the longitudinal polarisations (\ref{eq:longitudinal-polarization}), four factors $E^{1/2}$ from the external spinors (\ref{eq:external-spinors}), two factors $\kappa E$ from the vertex (\ref{graviton_gravitino_vertex_simp}) and one factor $1/E^2$ from the propagator (\ref{graviton_propagator}). For instance, the subamplitudes $t, u$ and $4f$ with helicities $(+, +, -, -)$ show the high-energy behaviour 
\begin{eqnarray}
\mathcal{M}^{+, +, -, -}_t &=& \frac{2\kappa^2E^6}{9m_{3/2}^4}(7+4\cos(\theta)-3\cos(2\theta))+\mathcal{O}\left(\frac{\kappa^2E^4}{m_{3/2}^2}\right),\\
\mathcal{M}^{+, +, -, -}_u &=& \frac{2\kappa^2E^6}{9m_{3/2}^4}(7-4\cos(\theta)-3\cos(2\theta)) +\mathcal{O}\left(\frac{\kappa^2E^4}{m_{3/2}^2}\right),
\end{eqnarray}

and
\begin{equation}
\mathcal{M}^{+, +, -, -}_{4f} = \frac{4\kappa^2E^6}{9m_{3/2}^4}(-7+3\cos(2\theta)) +\mathcal{O}\left(\frac{\kappa^2E^4}{m_{3/2}^2}\right).
\end{equation}

It follows that these three contributions cancel in the total amplitude, leaving the next order
\begin{equation}\label{eq:amplitude_++--}
\mathcal{M}^{+, +, -, -}_{\mathrm{grav}} = \frac{16\kappa^2E^4}{3m_{3/2}^2} +\mathcal{O}(\kappa^2E^2).
\end{equation}

This cancellation of the most divergent term also happens for other amplitudes, for instance
\begin{equation}\label{eq:amplitude_+++-}
\mathcal{M}^{+, +, +, -}_{\mathrm{grav}} = -\frac{4\kappa^2E^3}{3m_{3/2}}\sin(2\theta) +\mathcal{O}(\kappa^2m_{3/2}E),
\end{equation}

and
\begin{equation}\label{eq:amplitude_++++}
\mathcal{M}^{+, +, +, +}_{\mathrm{grav}} = \frac{16\kappa^2E^2}{3}\sin^2(\theta) +\mathcal{O}(\kappa^2m_{3/2}^2),
\end{equation}

where in the amplitude \eqref{eq:amplitude_+++-} there are terms behaving as $\kappa^2 E^5/m_{3/2}^3$ that cancel, while in \eqref{eq:amplitude_++++} there are terms behaving as $\kappa^2E^4/m_{3/2}^2$ that cancel as well. These cancellations are reminiscent of those we mentioned in Section \ref{sect:Higgs}, that lead to the scaling $\mathcal{M}_{\mathrm{gauge}}\sim E^2/v^2$ in (\ref{eq:amplitude-gauge}), instead of the expected $E^4/v^4$. They can therefore be understood similarly, using the Goldstino equivalence theorem: the high energy behaviour of the longitudinal gravitino amplitudes is captured by the corresponding Goldstino amplitudes, which scale at most like $\kappa^2E^4/m_{3/2}^2$ if we count the different contributions to the graphs as earlier. The difference in scaling for the different helicities can also be understood using the Goldstino theorem, which implies that (\ref{eq:amplitude_++--})-(\ref{eq:amplitude_++++}) correspond to scattering amplitudes of the Goldstino. Indeed, in Section \ref{sect:Polonyi}, we will see that their leading terms also cancel with the contributions coming from the scalar Goldstino superpartner of the Polonyi sector, leaving only the next order. Also, the amplitudes $(+, +, +, -)$ and $(+, +, +, +)$ have to vanish in the unbroken limit $m_{3/2}\rightarrow0$ where the Goldstino fermion is massless, because of chirality conservation at the graviton-Goldstino vertex. The power of $m_{3/2}$ in the $\mathcal{O}(\cdots)$ of (\ref{eq:amplitude_+++-})-(\ref{eq:amplitude_++++}) reflects the number of chirality violations when $m_{3/2}\rightarrow0$. With the new FI term, this limit cannot be taken so the argument does not apply, but the gravitational amplitudes remain the same.

\vfill\newpage\vfill

\section{Unitarity cutoff of the Polonyi model}
\label{sect:Polonyi}
The Polonyi model is the simplest 4-dimensional $\mathcal{N}=1$ supergravity where supersymmetry is broken spontaneously by the non-vanishing vacuum expectation value (vev) of a $F$-term. It is obtained by coupling pure supergravity to one chiral multiplet $\phi$\footnote{As usual, we denote by $\phi$ the chiral multiplet as well as its lowest complex scalar component.}, with a canonical K\"{a}hler potential and a superpotential of the form linear + constant as follows
\begin{equation}
    K(\phi,\bar{\phi})=\phi\bar{\phi} \quad \text{and} \quad W(\phi)=\mu M_{\mathrm{Pl}}(\phi+\beta),
\end{equation}
where $\mu$ and $\beta$ are two real constants with mass dimension $1$. The scalar potential of this model is given by
\begin{equation}\label{eq:scalar_potential}
    V=\mu^2\kappa^{-2}e^{\kappa^2\phi\bar{\phi}}\left(|1+\kappa^2\bar{\phi}(\phi+\beta)|^2-3\kappa^2|\phi+\beta|^2 \right).
\end{equation}
As can be easily checked by solving the equations $V=0$ and $\partial_{\bar{\phi}}V=0$, the model has a Minkowski vacuum solution with the scalar field at the real vev $\phi_0=1/(2\kappa)(\sqrt{3}-\kappa\beta)$, for two particular values of the parameter $\beta=\kappa^{-1}(\pm 2-\sqrt 3)$. The condition that this vacuum breaks supersymmetry, $\braket{\nabla_{\phi}W}_{\phi=\phi_0}\neq 0$, then selects $\beta=\kappa^{-1}(2-\sqrt 3)$. For this value, $\phi_0=\kappa^{-1}(\sqrt{3}-1)$, supersymmetry is broken by a non-vanishing vev of the $F$-term, and the Goldstino is proportional to the chiral fermion sitting in the chiral multiplet. In the unitary gauge, it is absorbed by the gravitino and eliminated from the spectrum, leading to a massive gravitino
\begin{equation}
m_{3/2}=\Braket{\kappa^2e^{\kappa^2K/2}W}_{\phi=\phi_0}=\mu e^{2-\sqrt 3}.
\end{equation}
Our aim is to compute the tree-level massive gravitino scattering amplitude $\psi\psi\rightarrow\psi\psi$ in this model. In the gravitational sector, the total amplitude for this process receives contributions from the $s, t, u$ channels with exchange of an off-shell graviton, as well as from the four gravitino contact term, as described in Section \ref{sect:mass_grav_scatt}. Because of the coupling with the chiral multiplet, in the Polonyi model, the total amplitude for the scattering $\psi\psi\rightarrow\psi\psi$ also receives contributions from the $s, t, u$ channels with exchange of an off-shell real scalar. These real scalars, called $x$ and $y$ in the following, are the real and imaginary fluctuations of $\phi$ above the  background $\phi_0=\kappa^{-1}(\sqrt{3}-1)$, and are defined in the following way
\begin{equation}\label{eq:scalar_fluctuations}
    \phi=\kappa^{-1}(\sqrt{3}-1)+\frac{x+iy}{\sqrt{2}},
\end{equation}
so that their kinetic terms are canonically normalised. Again, three ingredients are needed to compute the corresponding $s, t$ and $u$ graphs: the external polarisation of the gravitinos, whose computation has already been detailed in Section \ref{sect:mass_grav_scatt}, the propagators of the scalars $x$ and $y$, and the vertices $x\psi^2$ and $y\psi^2$.

\vfill\newpage\vfill

Expanding the scalar potential \eqref{eq:scalar_potential} to quadratic order in the fields, one finds that $x$ and $y$ have masses given by
\begin{equation}
    m_x^2=2\sqrt{3}m_{3/2}^2,\quad m_y^2=2(2-\sqrt{3})m_{3/2}^2,
\end{equation}
so that their propagators are given, in terms of the gravitino mass $m_{3/2}$ by
\begin{equation}\label{eq:scalar_propagators}
    P_{x}(k)=-\frac{i}{k^2+2\sqrt{3}m_{3/2}^2},\quad P_{y}(k)=-\frac{i}{k^2+2(2-\sqrt{3})m_{3/2}^2}.
\end{equation}

\vfill

The full Lagrangian of the above 4-dimensional $\mathcal{N}=1$ supergravity coupled to one chiral multiplet $\phi$, in the unitary gauge where the fermion is set to zero after SUSY breaking, reads
\begin{equation}\label{eq:full_lagrangian}
    e^{-1}\mathcal{L}=\frac{1}{2\kappa^2}\left(R(e)-\bar{\psi}_{\mu}\gamma^{\mu\rho\sigma}D_{\rho}\psi_{\sigma}\right)-g_{\phi\bar{\phi}}\partial_{\mu}\phi\partial^{\mu}\bar{\phi}-V+\mathcal{L}_{m_{3/2}}+\mathcal{L}_{4f},
\end{equation}
where the Lorentz-K\"{a}hler covariant derivative $D_{\rho}$ is given by
\begin{equation}
    D_{\rho}\equiv\partial_{\rho}+\frac{1}{4}\omega_{\rho}^{ab}(e)\gamma_{ab}-\frac{3}{2}i\mathcal{A}_{\rho}(\phi)\gamma_{*}.
\end{equation}
The spin connection $\omega_{\rho}^{ab}(e)$ and the K\"{a}hler connection $\mathcal{A}_{\rho}(\phi)$ are composite connections; the former has already been written in (\ref{eq:spin-connection}), while the latter is expressed as
\begin{equation}
\label{eq:kahler_connection}
\mathcal{A}_{\mu}(\phi)=\frac{i\kappa^2}{6}(\partial_{\mu}\phi\partial_{\phi}K-\partial_{\mu}\bar{\phi}\partial_{\bar{\phi}}K).
\end{equation}
The next term in (\ref{eq:full_lagrangian}) is $V=e^{\kappa^2 K}(-3\kappa^2 W\bar{W}+g^{\phi\bar{\phi}}\nabla_{\phi}W\bar{\nabla}_{\bar{\phi}}\bar{W})$, the scalar potential of $\mathcal{N}=1$, $D=4$ supergravity coupled to chiral multiplets. For the Polonyi model it is given in \eqref{eq:scalar_potential}. Then, $\mathcal{L}_{m_{3/2}}$ is the gravitino mass term,
\begin{equation}\label{eq:gravitino_mass_lagrangian}
    \mathcal{L}_{m_{3/2}}=\frac{\kappa^2}{2}e^{\kappa^2K/2}W\bar{\psi}_{\mu}P_R\gamma^{\mu\nu}\psi_{\nu}+h.c.,
\end{equation}
where $P_R$ is the right chiral projector $P_R=1/2(1-\gamma_*)$, with hermitian conjugate $P_L=1/2(1+\gamma_*)$. Finally, $\mathcal{L}_{4f}$ the four-gravitinos interaction term already written in \eqref{gravitino_quartic_interaction}.

\vfill

In order to find the $x\psi^2$ and $y\psi^2$ vertices, we expand the scalar field around its background according to \eqref{eq:scalar_fluctuations}. Scalar-gravitino interactions are contained in the gravitino mass term \eqref{eq:gravitino_mass_lagrangian}, and in the term involving the K\"{a}hler connection \eqref{eq:kahler_connection} because $\phi$ has a nonvanishing vev. The latter interaction involves only the derivative of $y$. Putting these together
\begin{eqnarray}
\mathcal{L}_{x\psi^2}&=&\frac{\kappa}{2}\sqrt{\frac{3}{2}}m_{3/2}~x~\bar{\psi}_{\mu}\gamma^{\mu\nu}\psi_{\nu},\\
\mathcal{L}_{y\psi^2}&=&-\frac{i\kappa}{2\sqrt{2}}m_{3/2}~y~\bar{\psi}_{\mu}\gamma_{*}\gamma^{\mu\nu}\psi_{\nu}-i\kappa\frac{(\sqrt{3}-1)\sqrt{2}}{8}~\partial_{\rho}y~\bar{\psi}_{\mu}\gamma^{\mu\rho\nu}\gamma_{*}\psi_{\nu}. 
\end{eqnarray}

\vfill

Obtaining the vertex from these Lagrangian is then straightforward. Taking into account the gravitino interchange, and using the same prescription as in (\ref{eq:graviton-gravitino-vertex}) for the factors of $i$
\begin{subequations}\label{eq:scalar_vertices}
\begin{align}
V_{x\psi^2}^{\mu\nu}&=i\sqrt{\frac{3}{2}}\kappa m_{3/2}\gamma^{\mu\nu},\\
V_{y\psi^2}^{\mu\nu}&=\frac{1}{\sqrt{2}}\kappa m_{3/2}\gamma_*\gamma^{\mu\nu}-i\kappa\frac{\sqrt{2}(\sqrt{3}-1)}{4}\gamma^{\mu\rho\nu}\gamma_*k_{\rho}.
\end{align}
\end{subequations}

As explained before equation (\ref{graviton_gravitino_vertex_simp}), we could use the fact that all gravitinos in our computation are on-shell to simplify this vertex with their equations of motion, e.g. $\gamma\cdot\psi=0$.

\vfill

Combining the polarisations of the external gravitinos \eqref{external_polarizations_gravitino}, the scalar propagators \eqref{eq:scalar_propagators}, the vertices \eqref{eq:scalar_vertices}, and the conventions depicted in Figure \ref{fig:diagrams}, we compute the amplitude contributions from scalars exchange. For given external gravitino helicities, the scalar contribution $\mathcal{M}_{\phi}^{\pm\pm\pm\pm}$ is the sum of the $s, t, u$ channel amplitudes, with either $x$ or $y$ propagating
\begin{equation}\label{eq:amplitude_scalar_sector}
    \mathcal{M}_{\phi}^{\pm\pm\pm\pm}=\sum_{\varphi=x,y}\sum_{j=s,t,u}\mathcal{M}_{j~(\varphi)}^{\pm\pm\pm\pm}.
\end{equation}

As in (\ref{eq:amplitude_++--})-(\ref{eq:amplitude_++++}) and explained in the end of Section \ref{sect:mass_grav_scatt}, the most divergent amplitude correspond to scattering with two  helicities $+1/2$ and two helicities $-1/2$, for instance $(+,+,-,-)$. The corresponding contribution from the scalar channels (\ref{eq:amplitude_scalar_sector}) is
\begin{equation}
\mathcal{M}^{+, +, -, -}_{\phi} = -\frac{16\kappa^2E^4}{3m_{3/2}^2} +\mathcal{O}(\kappa^2E^2),
\end{equation}
which exactly cancels the leading term of \eqref{eq:amplitude_++--}, leading to an amplitude
\begin{equation}
    \mathcal{M}_{\text{total}}^{+, +, -, -}=\mathcal{M}_{\mathrm{grav}}^{+, +, -, -}+\mathcal{M}_{\phi}^{+, +, -, -}\sim\kappa^2E^2.
\end{equation}
Applying the perturbative unitarity bound (\ref{eq:unitarity-bound}) one concludes that the unitarity cutoff of the theory lies at the Planck scale. This cancellation of the leading term between the gravitational and scalar channels also happens with other helicity assignments, for example
\begin{eqnarray}
\mathcal{M}^{+, +, +, -}_{\phi} &=& \frac{4\kappa^2E^3}{3m_{3/2}}\sin(2\theta) +\mathcal{O}(\kappa^2m_{3/2}E),\\
\mathcal{M}^{+, +, +, +}_{\phi} &=& -\frac{16\kappa^2E^2}{3}\sin^2(\theta) +\mathcal{O}(\kappa^2m_{3/2}^2),
\end{eqnarray}
whose most divergent terms exactly cancel with the ones of the amplitudes \eqref{eq:amplitude_+++-} and \eqref{eq:amplitude_++++}.\\

\vfill\newpage\vfill

At this point, some comments are in order. We recall that, in the Polonyi model under consideration, supersymmetry is broken by a non-vanishing $F$-term vev given by
\begin{equation}
    \Braket{F}=\Braket{\nabla_{\phi}W}=\sqrt{3}e^{-2+\sqrt{3}}~m_{3/2}/\kappa\sim m_{3/2}/\kappa,
\end{equation}
so the SUSY breaking scale is
\begin{equation}\label{eq:susy_breaking_scale}
    M_{\text{SUSY}}\equiv\sqrt{\Braket{F}}\sim \sqrt{m_{3/2}/\kappa}.
\end{equation}
If we considered only the gravitational channels, the unitarity criterion $|\mathcal{M}|<1$ of (\ref{eq:unitarity-bound}) applied to the most divergent amplitude (\ref{eq:amplitude_++--}) would lead to a cutoff at
\begin{equation}\label{eq:cutoff-grav}
    \Lambda\sim \sqrt{m_{3/2}/\kappa}\sim M_{\text{SUSY}}.
\end{equation}
The contributions from the scalar channels, by cancelling the leading term in $\mathcal{O}(\kappa^2E^4/m_{3/2}^2)$, uplift this cutoff from the supersymmetry breaking scale $M_{\text{SUSY}}$ to the Planck scale $M_{\mathrm{Pl}}$.

\vfill

This result can be seen as the simplest supersymmetric generalisation of the Standard Model mechanism reminded in Section \ref{sect:Higgs}. Depending on the vev of the Higgs field, two phases exist, where gauge symmetry is preserved and spontaneously broken. In the unbroken phase, the gauge bosons $W^{\pm}$ and $Z$ are massless, and their scattering amplitudes are well-behaved with respects to perturbative unitarity, i.e. they do not grow with energy. In the broken phase, the gauge bosons become massive. Since their longitudinal polarisations scale like $E/m$, it could lead to perturbative unitarity violation at high energy. This issue is automatically solved when considering the contributions from the Higgs channels, which exactly cancel the leading terms of the gauge channels, preserving unitarity in the broken phase as well. The situation is completely analogous for the Polonyi model considered here, replacing the Higgs field by the scalar partner of the Goldstino and the $W^\pm, Z$ gauge bosons by the gravitino $\psi_\mu$. 

\vfill

Both models have in common the existence of two phases, described by the same microscopic theory. It is therefore natural that perturbative unitarity is preserved in both phases up to the Planck scale, since this is a property of the same underlying theory. Having this in mind, it seems interesting to study a case where only one phase is well-defined. In the next Section, we consider $\mathcal{N}=1$, $D=4$ supergravity coupled to a vector multiplet, with a new type of Fayet-Iliopoulos term which does not require gauging the R-symmetry. To write this term we need a non-vanishing vev of the auxiliary field $D$, implying $D$-term supersymmetry breaking. When $\braket{D}\rightarrow 0$, the new FI term becomes singular, so there is no well-defined unbroken phase, and the situation of the cutoff is less clear than for the Polonyi model.


\vfill\newpage\vfill

\section{Unitarity cutoff of the new FI term}
\label{sect:cutoff_FI_term}
In $\mathcal{N}=1$, $D=4$ supergravity, a new Fayet-Iliopoulos term with ungauged R-symmetry has recently been introduced \cite{Cribiori:2017laj}. In the superconformal formalism\footnote{All along this section, we will use the superconformal formalism of matter-coupled $\mathcal{N}=1$, $D=4$ supergravity, following the conventions of \cite{Freedman:2012zz}. All notations used in this section are defined in Appendix \ref{sect:append_multiplet_calculus}.}, denoting $S_0=(s_0,P_L\Omega_0,F_0)$ and $\bar{S}_0=(\bar{s}_0,P_R\Omega_0,\bar{F}_0)$ the chiral and anti-chiral compensator multiplets, with (Weyl, Chiral) weights $(1,1)$ and $(1,-1)$ respectively, and $V = (A_\mu, \lambda, D)$ the vector multiplet in the Wess-Zumino gauge, this new FI Lagrangian is written as follows \cite{Cribiori:2017laj}
\begin{equation}\label{eq:FI_lagrangian_1}
\mathcal{L}_{\mathrm{FI}}=-\kappa^2\xi\left[S_0\bar{S}_0\frac{\mathcal{W}^2\overline{\mathcal{W}}^2}{T(\overline{\mathcal{W}}^2)\overline{T}(\mathcal{W}^2)}(V)_D \right]_D,
\end{equation}
where $\xi$ is a constant parameter of mass dimension two, $(V)_D$ is the real linear multiplet $(V)_D=(D,\slashed{\mathcal{D}}\lambda,0,\mathcal{D}^b\hat{F}_{ab},-\slashed{\mathcal{D}}\slashed{\mathcal{D}}\lambda,-\Box^C D)$ whose lowest component is the real auxiliary field of the vector multiplet, and $\mathcal{W}$ (resp. $\overline{\mathcal{W}}$) is its (anti-)chiral field strengths defined as
\begin{equation}\label{eq:field_strength_1}
\mathcal{W}^2=\frac{\bar{\lambda}P_L\lambda}{S_0^2}\quad\text{and}\quad\overline{\mathcal{W}}^2=\frac{\bar{\lambda} P_R\lambda}{\bar{S}_0^2}.
\end{equation}
Note that $(V)_D$ is the super-covariant derivative of $\mathcal{W}$. Then, $\bar{\lambda}P_L\lambda$ has weights $(3,3)$ and\footnote{As often in multiplet calculus, we use the same symbols to denote the multiplets and their lowest components.}
\begin{equation}\label{eq:lambda^2}
\bar{\lambda}P_L\lambda=\left(\bar{\lambda}P_L\lambda;\quad\sqrt 2P_L\left(-\frac{1}{2}\gamma\cdot\hat{F}+iD\right)\lambda;\quad2\bar{\lambda}P_L\slashed{\mathcal{D}}\lambda+\hat{F}^-\cdot\hat{F}^--D^2\right),    
\end{equation}
with the covariant field strength $\hat{F}_{ab}$ and the self-dual / anti self-dual tensors $\hat{F}_{ab}^{\pm}$ given by
\begin{equation}
\hat{F}_{ab}=e_a^{\mu}e_b^{\nu}(2\partial_{[\mu}A_{\nu]}+\bar{\psi}_{[\mu}\gamma_{\nu]}\lambda)\quad\text{and}\quad
\hat{F}_{ab}^{\pm}=\frac{1}{2}(\hat{F}_{ab}\pm\tilde{\hat{F}}_{ab})\quad\text{where}\quad
\tilde{\hat{F}}_{ab}=-\frac{i}{2}\epsilon_{abcd}\hat{F}^{cd}
\end{equation}
and
\begin{equation}
\mathcal{D}_{\mu}\lambda=\left(\partial_{\mu}-\frac{3}{2}b_{\mu}+\frac{1}{4}\omega_{\mu}^{ab}\gamma_{ab}-\frac{3}{2}i\gamma_*\mathcal{A}_{\mu}\right)\lambda-\left(\frac{1}{4}\gamma^{ab}\hat{F}_{ab}+\frac{1}{2}i\gamma_*D\right)\psi_{\mu}.
\end{equation}
The fields $b_{\mu}$ and $\mathcal{A}_{\mu}$ are the gauge fields corresponding respectively to dilatation and $T_R$ symmetry of the superconformal algebra. Finally, the chiral projection operator $T$ in (\ref{eq:FI_lagrangian_1}) acting on an anti-chiral multiplet $\bar{X}=(\bar{X},P_R\Omega,\bar{F})$ of weights $(1,-1)$ gives a chiral multiplet of weights $(2,2)$ defined as $T(\bar{X})=(\bar{F},\slashed{\mathcal{D}}P_R\Omega,\Box^C\bar{X})$. Applied on $\bar{\mathcal{W}^2}$ it yields 
\begin{equation}
T(\overline{\mathcal{W}}^2)=(C_T,P_L\Omega_T,F_T),
\end{equation}
with, in the gauge $\Omega_0=0$:
\begin{eqnarray}
\label{eq:C_T}
    C_T&=&\bar{s}_0^{-2}\left(2\bar{\lambda}P_R\slashed{\mathcal{D}}\lambda+\hat F^+\cdot\hat F^+-D^2\right)-2\bar{s}_0^{-3}\bar F_0\bar{\lambda}P_R\lambda,\\
    \label{eq:P_L_Omega_T}
    P_L\Omega_T&=&-\sqrt{2}\slashed{\mathcal{D}}\left(\bar{s}_0^{-2}P_R\left(\frac{1}{2}\gamma\cdot\hat F+iD\right)\lambda\right)\quad\text{and}\quad
    F_T=\square^C\left(\bar{s}_0^{-2}\bar{\lambda}P_R\lambda\right),
\end{eqnarray}

\vfill\newpage\vfill

and similarly for $\overline{T}(\mathcal{W}^2)$. An important point to notice is that if $D$ has a vanishing vacuum expectation value, then $C_T$ vanishes as well in the vacuum, making the new FI term singular. As a consequence, this term is non local in the limit of unbroken supersymmetry.

\vfill

The full Lagrangian of $\mathcal{N}=1$ supergravity coupled to an abelian $U(1)$ gauge multiplet and the new FI term is
\begin{equation}\label{eq:AdS_supergravity_plus_new_FI}
\mathcal{L}=-3\left[S_0\bar{S}_0\right]_D+\left[\kappa S_0^3 W_0\right]_F-\frac{1}{4g^2}\left[\bar{\lambda}P_L\lambda\right]_F+\mathcal{L}_{\mathrm{FI}}.
\end{equation}
where we included the possibility of a constant superpotential $W_0$ since the R-symmetry is not gauged. In components, and in the Poincaré gauge $s^0=\kappa^{-1}$ for the conformal symmetry,
\begin{eqnarray}
    -\frac{1}{4g^2}\left[\bar{\lambda}P_L\lambda\right]_F+\mathcal{L}_{FI}&=&-\frac{1}{4g^2}F_{\mu\nu}F^{\mu\nu}+\frac{1}{2g^2}D^2-\xi D+\mathcal{L}_\mathrm{\lambda}\nonumber\\
    &\sim&-\frac{1}{4}F_{\mu\nu}F^{\mu\nu}+\frac{1}{2}D^2-g\xi D+\mathcal{L}_\mathrm{\lambda}.
\end{eqnarray}
In the second line, we have canonically normalised the kinetic term of the $U(1)$ vector; also, $\mathcal{L}_\lambda$ denotes terms containing at least one gaugino $\lambda$. The gauge coupling $g$ can be absorbed into the FI paramter $\xi$, so we can set $g=1$ in the following. This leads to the standard equation of motion $D=\xi$. After eliminating the auxiliary fields of the gravitational multiplet as well as using their equations of motion, fixing the conformal symmetry with $s^0=\kappa^{-1}$, and in the unitary gauge where the Goldstino $\lambda$ is set to zero, the full Lagrangian \eqref{eq:AdS_supergravity_plus_new_FI} becomes
\begin{equation}\label{eq:lagrangian_FI_unitary}
e^{-1}\mathcal{L}=\frac{1}{2\kappa^2}\left(R-\bar{\psi}_{\mu}\gamma^{\mu\nu\rho}D_{\nu}\psi_{\rho}+m_{3/2}\bar{\psi}_{\mu}\gamma^{\mu\nu}\psi_{\nu}\right)-\frac{1}{4}F^{\mu\nu}F_{\mu\nu}-\left(-\frac{3}{\kappa^2}m_{3/2}^2+\frac{1}{2}\xi^2\right),
\end{equation}
with $m_{3/2}=W_0/2$. In order to use the results of Section \ref{sect:mass_grav_scatt} for the scattering of massive gravitinos in a Minkowski background, we need a vanishing cosmological constant. This can be obtained through a fine-tuning of the $\xi$ parameter
\begin{equation}
    \xi=\frac{\sqrt{6}m_{3/2}}{\kappa}.
\end{equation}
From the equation of motion $D=\xi$, we obtain the following supersymmetry breaking scale
\begin{equation}\label{eq:FI-Msusy}
    M_{\text{SUSY}}\equiv\sqrt{\Braket{D}}\sim \sqrt{\frac{m_{3/2}}{\kappa}}.
\end{equation}

\vfill

We are interested in the high energy behaviour of the supergravity theory defined by \eqref{eq:AdS_supergravity_plus_new_FI}. In particular we want to obtain its cutoff from the perturbative unitarity criterion (\ref{eq:unitarity-bound}). To this purpose, we first take a look at the energy scaling of the fermionic interaction terms contained in $\mathcal{L}_{FI}$, extending the dimensional analysis presented in \cite{Jang:2021vpb}. This will show that the cutoff is not lower than the supersymmetry breaking scale. We will then use the $2\rightarrow 2$ gravitino scattering amplitude of Section \ref{sect:mass_grav_scatt} to show that the cutoff indeed sits at $M_\mathrm{SUSY}$.

\vfill\newpage\vfill

The new FI Lagrangian \eqref{eq:FI_lagrangian_1} is the $D$-term density (\ref{eq:D_operation}) of a real multiplet built from the three chiral multiplets $S_0$, $W\equiv\bar{\lambda}P_L\lambda$ and $T\equiv T(\overline{\mathcal{W}}^2)$, their anti-chiral counterparts, and the real multiplet $(V)_D$. Its lowest component is $-\kappa^2\xi D f$, where $f$ is given by
\begin{equation}
    f(S_0, \bar S_0, W, \bar W, T, \bar T)\equiv (s_0\bar s_0)^{-1}\frac{(\bar{\lambda}P_L\lambda)(\bar{\lambda}P_R\lambda)}{C_TC_{\bar T}},
\end{equation}
with $C_T$ given in \eqref{eq:C_T}. One can now expand the $D$-term density in \eqref{eq:FI_lagrangian_1} following the multiplet calculus rules reminded in \eqref{eq:D_operation} and \eqref{eq:multiplet_composition_laws}. Apart from the bosonic FI term $-\xi D$, this yields a bunch of (non-renormalisable) fermionic interaction terms of the form
\begin{equation}\label{eq:general_scaling}
    \mathcal{L}_{FI}\supset -c\kappa^2\xi D^p (\partial_{s_0}^{m_0}\partial_W^{m_W}\partial_T^{m_T}f)\mathcal{O}_F^{(\delta)},
\end{equation}
where $c$ is a dimensionless constant and $p=0$ or $1$. The integers $m_0, m_W, m_T$ denote the number of derivatives of $f$ taken with respect to $S_0, W, T$ (see \eqref{eq:multiplet_composition_laws}). Such derivatives can also be taken with respect to $\bar S_0, \bar W, \bar T$, but for our purpose we can identify $S_0\sim \bar S_0$ and similarly for $W$ and $T$. With this in mind, we can simply write $f\sim s_0^{-2}W^2T^{-2}$. Finally, $\mathcal{O}_F^{(\delta)}$ is a fermionic operator of mass dimension $\delta$ fixed by homogeneity. Applying the derivatives on $f$ gives
\begin{eqnarray}
    \mathcal{L}_{FI}&\supset& -c\kappa^2\xi D^p s_0^{-2-m_0}W^{2-m_W}T^{-2-m_T}\mathcal{O}_F^{(\delta)}\\
    \label{eq:power_counting}
    &\supset& -c\xi \kappa^{m_0-2m_T} D^{-4-2m_T+p}\mathcal{O}_F^{(10+m_0+2m_T-2p)}.
\end{eqnarray}
In the second line, we used $s_0\sim \kappa^{-1}$ and $T\sim\kappa^2 D^2$ from \eqref{eq:C_T}; we also absorbed $W^{2-m_W}$ into $\mathcal{O}_F^{(\delta)}$ and fixed its mass dimension $\delta$ such that $\mathcal{L}_{FI}$ has mass dimension $4$ (remind that $D$ and $\xi$ have mass dimension $2$). At this point, there are factors of $\kappa$ in $\mathcal{O}_F$ coming from the gauge fixing of $s_0$. In order to conclude about the cutoff of these non-renormalisable operators, we should strip them from this $\kappa$ dependence. For this, we note that each derivative with respect to $T$ in \eqref{eq:general_scaling} is contracted with either $C_T$, $P_L\Omega_T$ or $F_T$ going into $\mathcal{O}_F$ in \eqref{eq:multiplet_composition_laws}. Then from \eqref{eq:C_T} and \eqref{eq:P_L_Omega_T}, we see that each of them comes with a factor $\kappa^2$. Similarly, each derivative with respect to $S_0$ brings either $s_0$, $P_L\Omega_0$ or $F_0$ in $\mathcal{O}_F$, which come with a $\kappa^{-1}$. There is no factor of $\kappa$ when taking derivatives with respect to $W$. In the end, we get
\begin{equation}\label{eq:power_counting_2}
    \mathcal{O}_F^{(\delta)}=\kappa^{-m_0+2m_T} \mathcal{O}_F^{'(\delta-m_0+2m_T)},
\end{equation}
where $\mathcal{O}_F^{'}$ is an operator containing only fields and derivatives, without factors of $\kappa$. Plugging \eqref{eq:power_counting_2} into \eqref{eq:power_counting}
\begin{equation}
\mathcal{L}_{FI}\supset-c\xi D^{-4-2m_T+p}\mathcal{O}_F^{'(10+4m_T-2p)}.
\end{equation}

\vfill\newpage\vfill

This is a term of the form $\mathcal{O}^{(4+n)}/\Lambda^{n}$, with $\Lambda$ a parameter of mass dimension $1$. In our case
\begin{equation}\label{eq:lower_bound_cutoff}
    \Lambda^{6+4m_T-2p}=\xi^{-1} D^{4+2m_T-p}=D^{3+2m_T-p}\quad\rightarrow\quad \Lambda_{\text{cutoff}}^{\text{FI}} \gtrsim \sqrt{D} = M_\mathrm{SUSY}
\end{equation}
using $D=\xi$ in the second equality. The scale $\Lambda$ is interpreted as the cutoff of the corresponding operator. It turns our that, on their own, \emph{all} the fermionic interactions contained in the new FI term have the same cutoff, which is the SUSY breaking scale from \eqref{eq:FI-Msusy}. This is a stronger result than the one presented in \cite{Jang:2021vpb}, where it was found that only some operators contributed to the lowest cutoff. When considering all these fermionic interactions together, however, the actual UV-cutoff obtained at the level of the Lagrangian could be higher than $M_\mathrm{SUSY}$ because of Fierz identities or potential cancellations in the amplitudes. To confirm that it sits indeed at the supersymmetry breaking scale, we will use the results of Section \ref{sect:mass_grav_scatt}.

\vfill

The Lagrangian \eqref{eq:AdS_supergravity_plus_new_FI} contains the gravitational Lagrangian \eqref{pure_sugra_action} considered in Section \ref{sect:mass_grav_scatt}. There, the most divergent $\psi\psi\rightarrow\psi\psi$ scattering amplitude is $\mathcal{M}^{+, +, -, -}_\mathrm{grav}$ in \eqref{eq:amplitude_++--}, corresponding to a cutoff $\Lambda^2\sim m_{3/2}/\kappa\sim M_\mathrm{SUSY}^2$ using \eqref{eq:FI-Msusy}. This cutoff could be uplifted, as in the Polonyi model, if the FI term opens new channels for the amplitude. The only degree of freedom that could propagate in such channels at tree level is the gauge boson of the vector multiplet which is the superpartner of the Goldstino. But for this, it needs to have an interaction term with two gravitinos. By construction of the new FI term, the $U(1)$ boson does not gauge the $R$-symmetry, so there is no minimal coupling $A\bar{\psi}\psi$ from the covariant derivative. It does not exclude the presence of non-minimal couplings coming from the new FI term. But after examining the component expansion of \eqref{eq:FI_lagrangian_1} given in \cite{Jang:2021vpb}, we see that all new interactions contain at least one gaugino $\lambda$ which vanishes in the unitary gauge. In other words, the new FI term does not contain any interaction term with one gauge field and two gravitinos, therefore it does not contribute to the $\psi\psi\rightarrow\psi\psi$ scattering amplitude at tree level.

\vfill

Consequently, the cancellation of the leading terms in $E$ between the gravitational and matter channels that we have highlighted in Section \ref{sect:Polonyi} for the Polonyi model does not happen for the new FI term. The most divergent amplitude remains \eqref{eq:amplitude_++--}, leading to the cutoff  
\begin{equation}
    \Lambda_{\text{cutoff}}^{\text{FI}}\sim \sqrt{\frac{m_{3/2}}{\kappa}}\sim M_{\text{SUSY}}.
\end{equation}
saturating the bound found in \eqref{eq:lower_bound_cutoff}. As already mentioned, the fundamental difference between the Polonyi and the new Fayet-Iliopoulos models is the absence of unbroken phase for the latter. In the Polonyi model, as well as in the Standard Model, it is natural that the ``soft'' UV behaviour of the amplitudes in the unbroken phase extends to the broken one, since the two are described by the same microscopic theory. With the new Fayet-Iliopoulos term, however, there was no reason for perturbative unitarity to be preserved up to the Planck scale, and we have indeed shown that it breaks down around the supersymmetry breaking scale.

\vfill\newpage\vfill

\section{Concluding remarks}

Within 4-dimensional $\mathcal{N}=1$ supergravity, we computed the $2\rightarrow2$ massive gravitino scattering amplitude at tree level, with external helicities $\pm1/2$. As in the case of longitudinal massive gauge boson amplitudes, we observed a cancellation of the leading terms between the different gravitational channels, which is a consequence of the Goldstino equivalence theorem. Then, in the Polonyi model of spontaneous sypersymmetry breaking, we found a second cancellation taking place, with the channels opened by the scalar partner of the Goldstino. This second cancellation brings the unitarity cutoff up to the Planck scale, in close analogy with the Standard Model, where the Higgs boson plays a crucial role in maintaining perturbative unitarity after electroweak symmetry breaking. Indeed, in both cases, the theory also contains an unbroken phase, where the longitudinal scattering amplitudes are well-behaved, so it is natural to expect a ``soft'' behaviour in the broken phase as well. When supersymmetry is broken by the new Fayet-Iliopoulos term, however, the unbroken phase is singular. In this case, we showed that the vector partner of the Goldstino does not contribute to the gravitino scattering amplitude at tree level and the cutoff remains at the supersymmetry breaking scale.\\

A cutoff at $M_\mathrm{SUSY}$ can also be encountered within non-linearly realised supersymmetry. In this framework, constraints are imposed to the multiplets. For example, in the case of one chiral multiplet, the Goldstino superfield $X$ is constrained to be nilpotent $X^2=0$. If we consider a model of supersymmetry breaking based on such a nilpotent multiplet, the cancellation we witnessed in Section \ref{sect:Polonyi} cannot happen, because the scalar partner of the Goldstino is removed by the nilpotency condition; thus the cutoff remains at $M_\mathrm{SUSY}$. In this case as well, the unbroken phase is not well defined, because degrees of freedom are missing in the multiplet. This is slightly different from what happens with the FI term: there, the vector partner of the Goldstino is still present, but decouples in the unitary gauge \eqref{eq:lagrangian_FI_unitary}. \\

\acknowledgments
F.R. was partially supported by the Cyprus R.I.F. Excellence hub/0421/0362 grant.

\vfill\newpage\vfill
\appendix
\section{Multiplet calculus}
\label{sect:append_multiplet_calculus}
This appendix summarises the notations and conventions for the multiplet calculus used in Section~\ref{sect:cutoff_FI_term}, and is based on \cite{Ferrara:2016een}. A general complex scalar multiplet is given by
\begin{equation}
\mathcal{C}=(\mathcal{C}, \mathcal{Z}, \mathcal{H}, \mathcal{K}, \mathcal{B}_{\mu}, \Lambda,\mathcal{D}),
\end{equation}
where $\mathcal{C}$, $\mathcal{H}$, $\mathcal{K}$ and $\mathcal{D}$ are complex scalars, $\mathcal{Z}$ and $\Lambda$ are Dirac fermions, and $\mathcal{B}_{\mu}$ is a Lorentz vector. 
A chiral multiplet is a complex multiplet having $P_R\mathcal{Z}=0$, $\mathcal{K}=0$, $\mathcal{B}_{\mu}=i\mathcal{D}_{\mu}\mathcal{C}$, $\Lambda=0$ and $\mathcal{D}=0$. Renaming $\mathcal{C}=Z$, it is written, in a seven-components notation, as
\begin{equation}
(Z, -i\sqrt{2}P_L\chi, -2F, 0, i\mathcal{D}_{\mu}Z, 0, 0),
\end{equation}
and similarly for its anti-chiral counterpart
\begin{equation}
(\bar{Z}, i\sqrt{2}P_R\chi, 0, -2\bar{F}, -i\mathcal{D}_{\mu}\bar{Z}, 0, 0).
\end{equation}
The (anti-)chiral multiplets can also be written in a three-components notation
\begin{equation}
(Z, P_L\chi, F)\quad\text{and}\quad(\bar{Z}, P_R\chi, \bar{F}).
\end{equation}

A real multiplet is a complex multiplet whose lowest component $\mathcal{C}=C$ is real. This implies that $\mathcal{Z}=\zeta, \Lambda=\lambda$ are Majorana spinors and $\mathcal{B}_{\mu}=B_{\mu}, \mathcal{D}=D$ are real, while $\mathcal{K}=\bar{\mathcal{H}}$ are still complex. A real multiplet is thus written, in a six-components notation, as
\begin{equation}
(C, \zeta, \mathcal{H}, B_{\mu}, \lambda, D).
\end{equation}   

The $F$-term density operator $[~]_F$ acts on a chiral multiplet of (Weyl, Chiral) weights $(3,3)$ as:
\begin{equation}
[~]_F:(Z,P_L\chi,F)\rightarrow[Z]_F\equiv e~\left[F+\frac{1}{\sqrt 2}\bar\psi_{\mu}\gamma^{\mu}P_L\chi+\frac{1}{2}Z\bar\psi_{\mu}\gamma^{\mu\nu}P_R\psi_{\nu}\right]+h.c.
\end{equation}
The $D$-term density operator $[~]_D$ acts on a real multiplet of weights $(2,0)$ according to:
\begin{eqnarray}\label{eq:D_operation}
[~]_D:(C, \zeta, \mathcal{H}, B_{\mu}, \lambda, D)\rightarrow[C]_D\equiv \frac{e}{2}\left[D-\frac{1}{2}\bar\psi_{\mu}\gamma^{\mu}i\gamma_*\lambda-\frac{1}{3}CR(\omega)\right.\nonumber\\
\quad\left.+\frac{1}{6}\left(C\bar\psi_{\mu}\gamma^{\mu\rho\sigma}-i\bar\zeta\gamma^{\rho\sigma}\gamma_*\right)R'_{\rho\sigma}(Q)\right.
\left.+\frac{1}{4}\epsilon^{abcd}\bar\psi_a\gamma_b\psi_c\left(B_d-\frac{1}{2}\bar\psi_d\zeta\right)\right],
\end{eqnarray}
where $R(\omega)$ and $R'_{\rho\sigma}(Q)$ are respectively the graviton and gravitino curvatures. Both operators are used to build superconformal invariant actions from chiral and real multiplets, respectively, according to $S_F=\int d^4x ~[Z]_F$ and $S_D=\int d^4x ~[C]_D$.

\vfill\newpage\vfill

Given a set of complex multiplets $\mathcal{C}^i=(\mathcal{C}^i, \mathcal{Z}^i, \mathcal{H}^i, \mathcal{K}^i, \mathcal{B}_{\mu}^i, \Lambda^i,\mathcal{D}^i)$, one can build another complex multiplet $\mathcal{C}=(\mathcal{C}, \mathcal{Z}, \mathcal{H}, \mathcal{K}, \mathcal{B}_{\mu}, \Lambda,\mathcal{D})$ whose lowest component is given by an arbitrary function $f$ of the first components of the $\mathcal{C}^i$, namely $\mathcal{C}=f(\mathcal{C}^i)$. The higher components of $\mathcal{C}$ are then given by:
\begin{eqnarray}\label{eq:multiplet_composition_laws}
\mathcal{Z}&=&f_i\mathcal{Z}^i, \nonumber \\
\mathcal{H}&=&f_i\mathcal{H}^i-\frac{1}{2}f_{ij}\mathcal{\bar{Z}}^iP_L\mathcal{Z}^j, \nonumber \\
\mathcal{K}&=&f_i\mathcal{K}^i-\frac{1}{2}f_{ij}\mathcal{\bar{Z}}^iP_R\mathcal{Z}^j, \nonumber \\
\mathcal{B}_{\mu}&=&f_i\mathcal{B}_{\mu}^i+\frac{1}{2}if_{ij}\mathcal{\bar{Z}}^iP_L\gamma_{\mu}\mathcal{Z}^j, \\
\Lambda&=&f_i\Lambda^i+\frac{1}{2}f_{ij}\left(i\gamma_{*}\slashed{\mathcal{B}}^i+P_L\mathcal{K}^i+P_R\mathcal{H}^i-\slashed{\mathcal D}\mathcal{C}^i \right)\mathcal{Z}^j-\frac{1}{4}f_{ijk}\mathcal{Z}^i \mathcal{\bar{Z}}^j \mathcal{Z}^k, \nonumber \\
\mathcal{D}&=&f_i\mathcal{D}^i+\frac{1}{2}f_{ij}\left(\mathcal{K}^i\mathcal{H}^j-\mathcal{B}^i\cdot\mathcal{B}^j-\mathcal{D}\mathcal{C}^i\cdot\mathcal{D}\mathcal{C}^j-2\bar{\Lambda}^i\mathcal{Z}^j-\mathcal{\bar{Z}}^i\slashed{\mathcal{D}}\mathcal{Z}^j \right), \nonumber \\ &-&\frac{1}{4}f_{ijk}\mathcal{\bar{Z}}^i\left(i\gamma_{*}\slashed{\mathcal{B}}^j+P_L\mathcal{K}^j+P_R\mathcal{H}^j \right)\mathcal{Z}^k +\frac{1}{8}f_{ijkl}\mathcal{\bar{Z}}^iP_L\mathcal{Z}^j\mathcal{\bar{Z}}^k P_R\mathcal{Z}^l,\nonumber
\end{eqnarray}
with $f_i\equiv\partial f/\partial \mathcal{C}^i$ and so on for higher order derivatives; summation over repeated indices is understood. The bar on spinors are the Majorana conjugate defined by $\bar\psi=\psi^TC$, with $C$ the usual charge conjugation matrix satisfying $\gamma_{\mu}^T=-C\gamma_{\mu}C^{-1}$. \\

\bibliography{bibliographie}

\begingroup\raggedright\begin{thebibliography}{20}
\expandafter\ifx\csname natexlab\endcsname\relax\def\natexlab#1{#1}\fi

\bibitem[Polonyi(1977)]{Polonyi:1977pj}
J.~Polonyi, ``{Generalization of the Massive Scalar Multiplet Coupling to the
  Supergravity}'', 1977.

\bibitem[Fayet and Iliopoulos(1974)]{Fayet:1974jb}
P.~Fayet and J.~Iliopoulos, ``{Spontaneously Broken Supergauge Symmetries and
  Goldstone Spinors}'', {\em Phys. Lett. B} {\bfseries 51} (1974) 461--464.

\bibitem[Freedman(1977)]{Freedman:1976uk}
D.~Z. Freedman, ``{Supergravity with Axial Gauge Invariance}'', {\em Phys. Rev.
  D} {\bfseries 15} (1977) 1173.

\bibitem[Barbieri et~al.(1982)Barbieri, Ferrara, Nanopoulos, and
  Stelle]{Barbieri:1982ac}
R.~Barbieri, S.~Ferrara, D.~V. Nanopoulos, and K.~S. Stelle, ``{Supergravity, R
  Invariance and Spontaneous Supersymmetry Breaking}'', {\em Phys. Lett. B}
  {\bfseries 113} (1982) 219--222.

\bibitem[Komargodski and Seiberg(2009)]{Komargodski:2009pc}
Z.~Komargodski and N.~Seiberg, ``{Comments on the Fayet-Iliopoulos Term in
  Field Theory and Supergravity}'', {\em JHEP} {\bfseries 06} (2009) 007,
  \href{http://xxx.lanl.gov/abs/0904.1159}{{\ttfamily arXiv:0904.1159}}.

\bibitem[Dienes and Thomas(2010)]{Dienes:2009td}
K.~R. Dienes and B.~Thomas, ``{On the Inconsistency of Fayet-Iliopoulos Terms
  in Supergravity Theories}'', {\em Phys. Rev. D} {\bfseries 81} (2010) 065023,
   \href{http://xxx.lanl.gov/abs/0911.0677}{{\ttfamily arXiv:0911.0677}}.

\bibitem[Cribiori et~al.(2018)Cribiori, Farakos, Tournoy, and van
  Proeyen]{Cribiori:2017laj}
N.~Cribiori, F.~Farakos, M.~Tournoy, and A.~van Proeyen, ``{Fayet-Iliopoulos
  terms in supergravity without gauged R-symmetry}'', {\em JHEP} {\bfseries 04}
  (2018) 032,  \href{http://xxx.lanl.gov/abs/1712.08601}{{\ttfamily
  arXiv:1712.08601}}.

\bibitem[Cribiori et~al.(2019)Cribiori, Farakos, and Tournoy]{Cribiori:2018dlc}
N.~Cribiori, F.~Farakos, and M.~Tournoy, ``{Supersymmetric Born-Infeld actions
  and new Fayet-Iliopoulos terms}'', {\em JHEP} {\bfseries 03} (2019) 050,
  \href{http://xxx.lanl.gov/abs/1811.08424}{{\ttfamily arXiv:1811.08424}}.

\bibitem[Antoniadis et~al.(2018{\natexlab{a}})Antoniadis, Chatrabhuti, Isono,
  and Knoops]{Antoniadis:2018cpq}
I.~Antoniadis, A.~Chatrabhuti, H.~Isono, and R.~Knoops,
  ``{Fayet\textendash{}Iliopoulos terms in supergravity and D-term
  inflation}'', {\em Eur. Phys. J. C} {\bfseries 78} (2018){\natexlab{a}},
  no.~5, 366,  \href{http://xxx.lanl.gov/abs/1803.03817}{{\ttfamily
  arXiv:1803.03817}}.

\bibitem[Antoniadis et~al.(2018{\natexlab{b}})Antoniadis, Chatrabhuti, Isono,
  and Knoops]{Antoniadis:2018oeh}
I.~Antoniadis, A.~Chatrabhuti, H.~Isono, and R.~Knoops, ``{The cosmological
  constant in Supergravity}'', {\em Eur. Phys. J. C} {\bfseries 78}
  (2018){\natexlab{b}}, no.~9, 718,
  \href{http://xxx.lanl.gov/abs/1805.00852}{{\ttfamily arXiv:1805.00852}}.

\bibitem[Antoniadis and Rondeau(2020)]{Antoniadis:2019nwz}
I.~Antoniadis and F.~Rondeau, ``{New K\"ahler invariant
  Fayet\textendash{}Iliopoulos terms in supergravity and cosmological
  applications}'', {\em Eur. Phys. J. C} {\bfseries 80} (2020), no.~4, 346,
  \href{http://xxx.lanl.gov/abs/1912.08117}{{\ttfamily arXiv:1912.08117}}.

\bibitem[Jang and Porrati(2021)]{Jang:2021fce}
H.~Jang and M.~Porrati, ``{Inflation, gravity mediated supersymmetry breaking,
  and de Sitter vacua in supergravity with a K\"ahler-invariant
  Fayet-Iliopoulos term}'', {\em Phys. Rev. D} {\bfseries 103} (2021), no.~10,
  105006,  \href{http://xxx.lanl.gov/abs/2102.11358}{{\ttfamily
  arXiv:2102.11358}}.

\bibitem[Jang and Porrati(2022)]{Jang:2022sql}
H.~Jang and M.~Porrati, ``{Realization of slow-roll inflation and the MSSM in
  supergravity theories with new Fayet-Iliopoulos terms}'', {\em Phys. Rev. D}
  {\bfseries 106} (2022), no.~4, 045024,
  \href{http://xxx.lanl.gov/abs/2207.01889}{{\ttfamily arXiv:2207.01889}}.

\bibitem[Jang and Porrati(2021)]{Jang:2021vpb}
H.~Jang and M.~Porrati, ``{Component actions of liberated $ \mathcal{N} $ = 1
  supergravity and new Fayet-Iliopoulos terms in superconformal tensor
  calculus}'', {\em JHEP} {\bfseries 11} (2021) 075,
  \href{http://xxx.lanl.gov/abs/2108.04469}{{\ttfamily arXiv:2108.04469}}.

\bibitem[Schwartz(2014)]{Schwartz:2014sze}
M.~D. Schwartz, ``{Quantum Field Theory and the Standard Model}'', Cambridge
  University Press, 3 2014.

\bibitem[Freedman and Van~Proeyen(2012)]{Freedman:2012zz}
D.~Z. Freedman and A.~Van~Proeyen, ``{Supergravity}'', Cambridge Univ. Press,
  Cambridge, UK, 5 2012.

\bibitem[Cremmer et~al.(1983)Cremmer, Ferrara, Girardello, and
  Van~Proeyen]{Cremmer:1982en}
E.~Cremmer, S.~Ferrara, L.~Girardello, and A.~Van~Proeyen, ``{Yang-Mills
  Theories with Local Supersymmetry: Lagrangian, Transformation Laws and
  SuperHiggs Effect}'', {\em Nucl. Phys. B} {\bfseries 212} (1983) 413.

\bibitem[Bjerrum-Bohr and Engelund(2010)]{Bjerrum-Bohr:2010eeh}
N.~E.~J. Bjerrum-Bohr and O.~T. Engelund, ``{Gravitino Interactions from
  Yang-Mills Theory}'', {\em Phys. Rev. D} {\bfseries 81} (2010) 105009,
  \href{http://xxx.lanl.gov/abs/1002.2279}{{\ttfamily arXiv:1002.2279}}.

\bibitem[Moroi(1995)]{Moroi:1995fs}
T.~Moroi, ``{Effects of the gravitino on the inflationary universe}'', ph.d
  thesis, 3 1995.
\newblock  \href{http://xxx.lanl.gov/abs/hep-ph/9503210}{{\ttfamily
  hep-ph/9503210}}.

\bibitem[Ferrara et~al.(2016)Ferrara, Kallosh, Van~Proeyen, and
  Wrase]{Ferrara:2016een}
S.~Ferrara, R.~Kallosh, A.~Van~Proeyen, and T.~Wrase, ``{Linear Versus
  Non-linear Supersymmetry, in General}'', {\em JHEP} {\bfseries 04} (2016)
  065,  \href{http://xxx.lanl.gov/abs/1603.02653}{{\ttfamily
  arXiv:1603.02653}}.

\end{thebibliography}\endgroup

\end{document}